\pdfoutput=1
\documentclass[aip,reprint]{revtex4-1}
\usepackage{textcomp}
\usepackage{bm}%
\usepackage{graphicx}
\usepackage{makecell}
\usepackage{array}
\usepackage{amsmath}
\usepackage{color}
\draft 

\begin{document}

\title
{Deconstructing temperature gradients across fluid interfaces: the structural origin of the thermal resistance of liquid-vapor interfaces}


\author{Jordan Muscatello}
\affiliation
{Department of Chemical Engineering, Imperial College London,
SW7 2AZ, London, United Kingdom}
\author{Enrique Chac\'{o}n}
\affiliation
{Instituto de Ciencia de Materiales de Madrid, CSIC, Sor Juana In\'es de la Cruz 3.3, E-28049, Madrid, Spain}
\author{Pedro Tarazona}
\affiliation
{Departamento de F\'isica Te\'orica de la Materia Condensada, IFIMAC Condensed
Matter Physics Center, Universidad Aut\'onoma de Madrid, Madrid 28049, Spain}
\author{Fernando Bresme}\email{f.bresme@imperial.ac.uk}
\affiliation
{Department of Chemistry, Imperial College London,
SW7 2AZ, London, United Kingdom}



\date{\today}

\begin{abstract}
The interfacial thermal resistance 
determines 
condensation-evaporation processes and thermal transport across material-fluid 
interfaces.
Despite its importance in transport processes, the interfacial structure 
responsible for the thermal 
resistance is still unknown.
By combining 
non-equilibrium molecular dynamics simulations
and interfacial analyses that remove the interfacial thermal fluctuations  
we show that the 
thermal resistance of liquid-vapor interfaces is connected to 
a low density fluid layer that is adsorbed at the liquid surface. This thermal 
resistance layer (TRL) defines the boundary where the thermal transport mechanism
changes from that of gases (ballistic) to that characteristic of dense liquids, dominated
by frequent particle collisions involving very short mean free paths.
We show that the thermal conductance is 
proportional to the number of atoms adsorbed in the TRL, and hence we explain 
the 
structural origin of the thermal resistance in liquid-vapor interfaces.
\end{abstract}
\maketitle

The interfacial thermal resistance describes the abrupt change  
of temperature, $\Delta T$, at the surface between two distinct bulk phases, 
when a heat flux $J_q$ goes across it.  
The effect was originally postulated for liquid helium/solid interfaces, 
as a consequence of the sharp change in the nature 
of the heat carriers, and it was first measured by Kapitza in 1941
\cite{kapitza1941,swartz1989} in terms of what is known as the Kapitza resistance
$R_K=\Delta T/J_q$, or the interfacial conductance 
$G_K=1/R_K$. 
In a bulk phase, the thermal conductivity $\lambda$ gives the ratio between 
the temperature gradient and the heat current, so that 
the different bulk conductivities 
produce different slopes in the temperature profile $T(z)$ at the two sides of the
interface. At the interface, the temperature profile features a 
sharp `jump', $\Delta T = J_q/ G_K$, reflecting the physical mechanism
associated to energy transfer between heat carriers in the different phases in contact. Hence, 
$R_K$ or $G_K$ provide insight into the molecular structure of the interface and the 
microscopic mechanism determining energy transport. 

Since the seminal work by Kapitza, the
concept of interfacial thermal resistance has been extended to describe
energy flux across many other interfaces and to identify the relevant transport channels for heat transfer. 
Understanding the physical origin of the thermal resistance would provide fundamental
information on the coupling between the heat carriers at the two sides 
of an interface. This problem is of significant interest in many current technologies, {\it e.g.} in microelectronics, the
management of energy transport in the form of heat is of utmost importance. 
Further, with the advent of ever more advanced experimental techniques,
the increasing spatial and temporal resolution 
to characterize thermal transport is approaching the atomistic length scales characteristic
of nanometric devices~\cite{Cui1192,capinski1999}.

Thermal conduction across fluid-fluid interfaces sets also important fundamental 
and applied challenges for the study of 
condensation-evaporation processes~\cite{Caputa2011}, 
thermal resistance of nanomaterials~\cite{C2CP43771F,wilson2002,Cahill2014,Tascini} 
and biological interfaces~\cite{LeitnerStraub,Yu2005,Lervik2010}. 
The 
thermal resistance
can be quite significant when the phases in contact involve appreciable
differences in density, {\it e.g.}, in a liquid-vapor interface
away from the critical region the large density difference
between the liquid ($l$) and the ($g$) phases ($\rho_{l}/\rho_{g} \sim 10^2-10^3$) involves a
qualitative change in the mechanism of heat transport. In liquids the mechanism is dominated by molecular collisions of atoms/molecules inside the cage formed by the nearest neighbors, without the need of mass diffusion. In low density gases the thermal transport involves 
long ballistic flights and long mean free paths, {\it e.g.} ten to hundred molecular diameters 
near triple point conditions. 
Under a given heat flow, the bulk thermal conductivities are very different, $\lambda_{l}\gg \lambda_{g}$ 
resulting in important changes in the 
slope of the temperature profile across the interface~
\cite{Rosjorde2000,Simon2004,Jackson2016}. Theoretical studies demonstrate
that kinetic theory offers an accurate approach to quantify
the thermal conductance of simple fluids and good agreement with 
simulations has been reported~\cite{kjesltrupbook}.

At the sharpest molecular scale employed in computer simulations, 
the temperature profile $T(z)$ may be consistently obtained either 
from the mean kinetic energy or the forces~\cite{Jackson2016} on atoms or molecules 
located at 
position $z$. 
The slope of $T(z)$ at the vapor and liquid phases 
is determined by  
the thermal conductivity of each phase, $\lambda_g$ and  $\lambda_l$. 
Kapitza's mesoscopic view considers the approximate description of $T(z)$ as
straight lines with slopes $J_q/\lambda_g$ and  $J_q/\lambda_l$, with a temperature `jump' $\Delta T=J_q/G_K$ at position $z_K$ (see Supporting Information).
Intriguingly, previous theoretical studies indicate 
that $z_K$ is  
shifted towards the vapor phase. 
The molecular interpretation of $z_K$ and $G_K$ has been elusive because fluid interfaces, 
liquid/vapor or liquid/liquid, 
feature 
capillary waves that {\it blur out} the molecular structure of the interface~\cite{RowlinsonBook}. This results
in 
a broadening of the density ($\rho(z)$) and temperature ($T(z)$) profiles, 
which 
eliminates 
important structural details of the interface.
Our hypothesis is that by removing the thermal fluctuations we can 
resolve the interfacial structure and hence identify the key regions 
defining the interfacial thermal resistance. To illustrate our idea, we study the liquid-vapor interface of a simple fluid.

%


\begin{figure*}[!ht]
\includegraphics[width=0.8\linewidth]{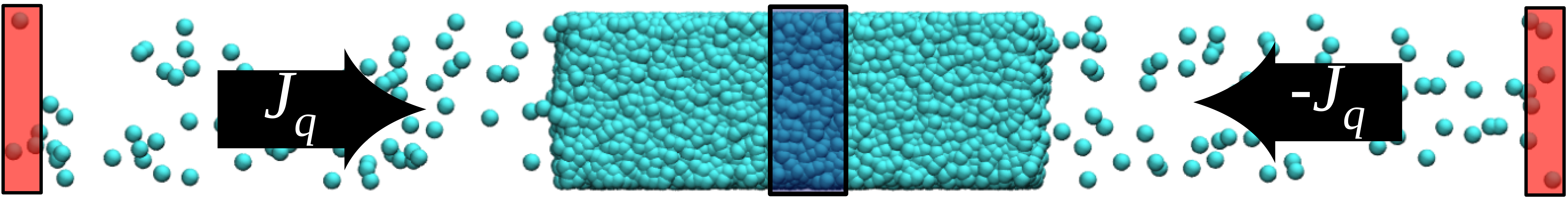}
 \caption{\label{fig:system}Schematic diagram showing the periodic
 simulation cell containing a liquid-vapor interface. The liquid phase is
 in the center of the box. The spheres represent the  atoms. Kinetic energy
 is extracted in the centre of the liquid phase (blue highlighted region) and
 added in the center of the gas phase (red highlighted regions) at a constant
 rate. 
 The arrows indicate the direction of the
 resultant energy flux. It should be noted that the actual simulation cells
 used in the simulations are significantly more elongated in the $z$-direction
 than in this snapshot.}
\end{figure*}
 
Stillinger introduced the concept of {\it intrinsic surface} (IS), a time dependent
surface that can be used to describe the interfacial structure by removing 
capillary wave effects~\cite{Stillinger1982}. 
%
%
The intrinsic
sampling method (ISM)~\cite{chacon} allows to identify the IS, making it possible  
to resolve the molecular structure of the interface~\cite{chacon,bresme-chacon2008}.
In this work 
we apply the ISM to a liquid-vapor interface
under an applied heat flux. By doing so we are able to identify the structural origin of the thermal resistance.


The simulations were performed on a system of Lennard-Jones particles at varying temperatures along the coexistence line.
Reduced units are used for all the quantities
throughout the  paper, unless otherwise indicated.
The heat flux was induced by 
creating a region of energy
extraction (cold region) in the centre of the liquid slab and by adding energy at
the edges of the simulation box, in the 
vapor phase (hot region) (see Figure~1). The heat flux 
is given by,
\begin{equation}
 J_{q}=\frac{\pm \dot{Q}}{2L_{x}L_{y}}
\end{equation}
Where $\dot{Q}$ is the rate of kinetic energy added(+)/withdrawn(-) from the hot/cold regions, and $L_x L_y$ is the cross sectional area of the simulation box (see Supplementary Information and references~\cite{LAMMPSmanual,hafskjold1993}). We used $\dot{Q}=5$ or $10$ for system I and II, III, 
respectively (see Figure~2). 
For system I and II 
the heat flux corresponds to $3.3\times10^7\text{ Wm}^{-2}$ in SI units. 




Our simulations reach a stationary state with no net mass flux, which allows the decoupling of heat and mass transfer processes and hence the computation of the thermal resistance from the temperature profiles (see ref.~\cite{KjelstrupBedeaux} for a discussion of coupled processes). Lack of thermalization may appear in the Knudsen layer in systems featuring strong evaporation. This effect is not relevant in our case, due to the lack of mass flux. This is supported by the similarity of temperature profiles obtained for
systems featuring very different vapor densities, and by earlier work that showed the consistency of the local temperatures computed using kinetic and configurational approaches~\cite{Jackson2016}. 

In the capillary wave theory formalism the intrinsic surface is defined as
\begin{equation}
 z=\xi(\mathbf{R}, q_u)\;,\;\mathbf{R}=(x,y)
\end{equation}
where the wavevector cutoff, $q_u$, determines the level of resolution of the surface. The intrinsic surface can be expressed in terms of a Fourier series,
\begin{equation}
 \xi(\mathbf{R},q_u) = \hat{\xi}_0 + \sum_{0<|\mathbf{q}|<q_u} \hat{\xi}_q
e^{i\mathbf{q}\cdot\mathbf{R}}.
\end{equation}
The ISM identifies the intrinsic surface via a percolation method,
and uses an iterative procedure to calculate the Fourier coefficients associated with each wavevector in the
expansion of the intrinsic surface. This is performed for various occupancy values, $n_S=N_S/A_0$, where $N_S$ is the number of atoms at the intrinsic
surface and $A_0$ is the cross sectional area of the interface.
We chose in this work $n_s = 1.0$ for systems I and II 
and $n_s = 0.7$ for system III. 
We find that these parameters provide a good
resolution of the interfacial structure (see the SI for other occupancies).
The intrinsic density and temperature profiles were computed with respect to the
intrinsic surface $\xi(\mathbf{R})$ defined above.

\begin{figure}[!ht]
\centering
\begin{tabular}{cc}
\includegraphics[width=0.5\linewidth]{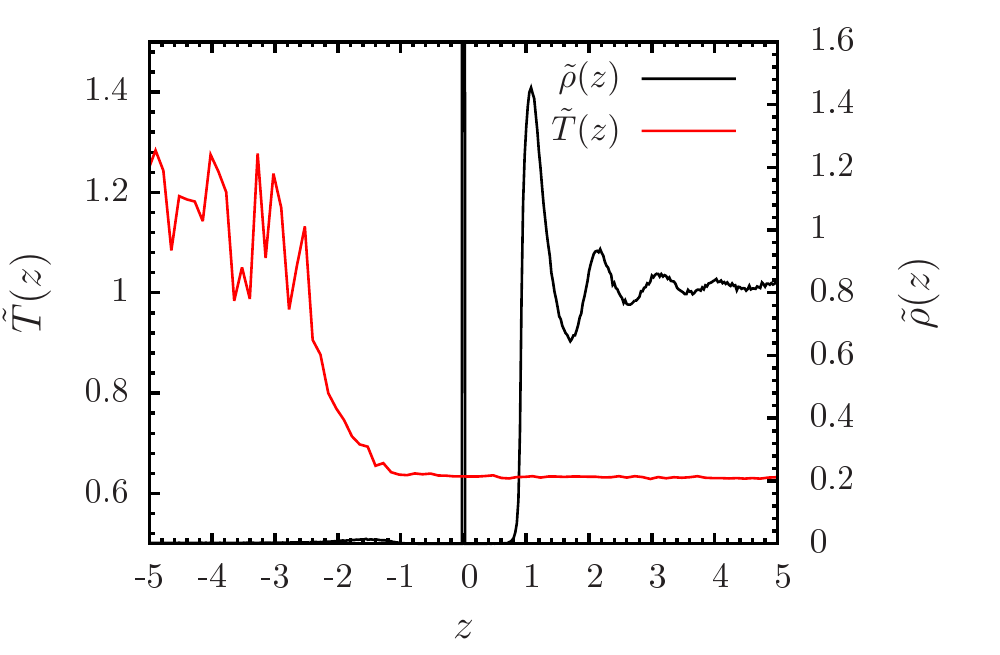} &
\includegraphics[width=0.5\linewidth]{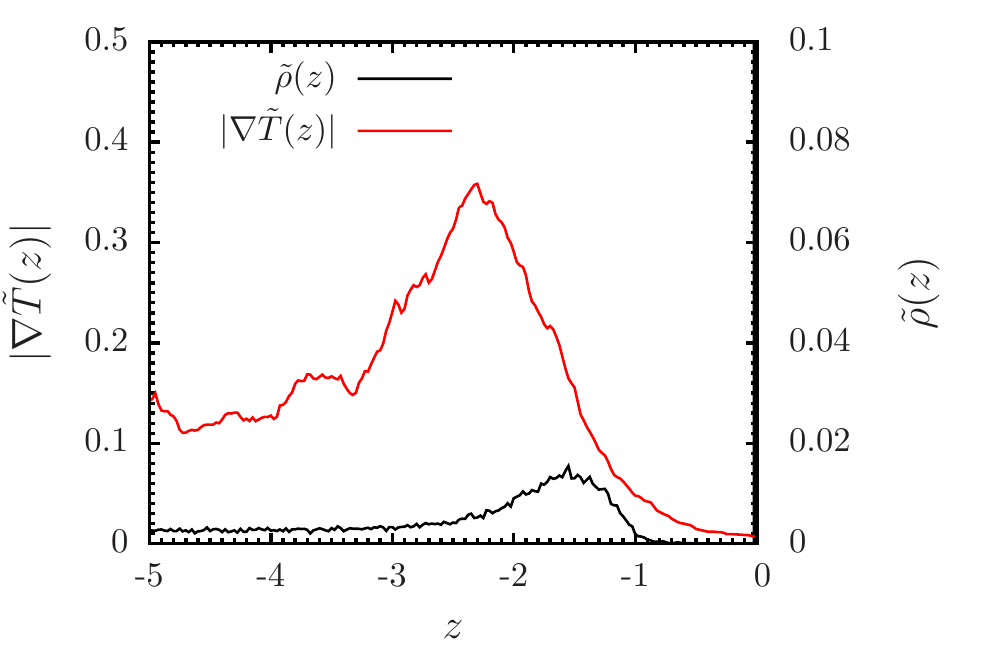} \\
\includegraphics[width=0.5\linewidth]{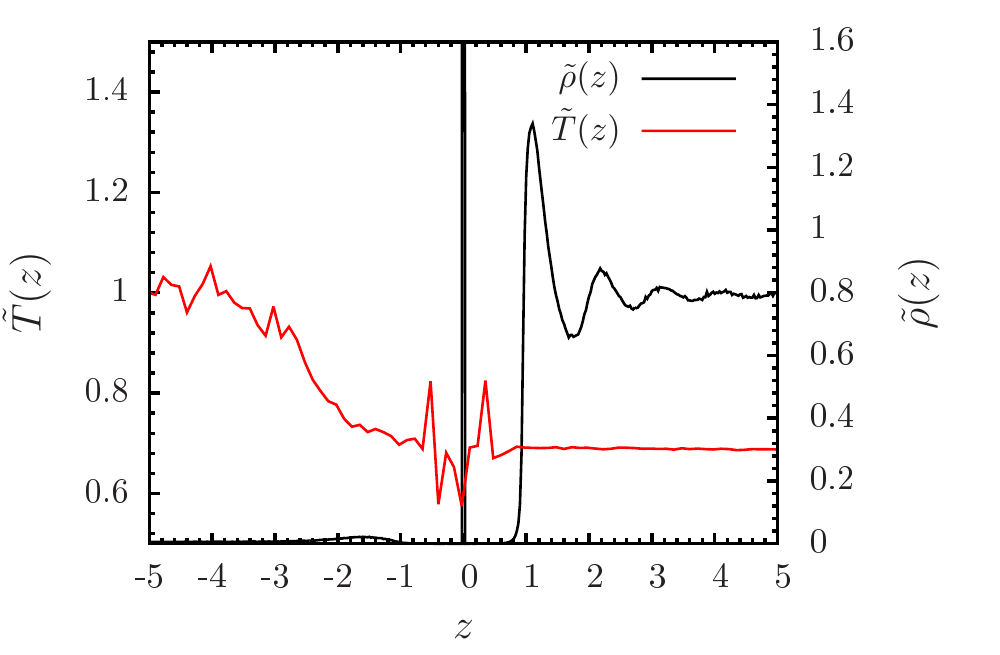}  &
\includegraphics[width=0.5\linewidth]{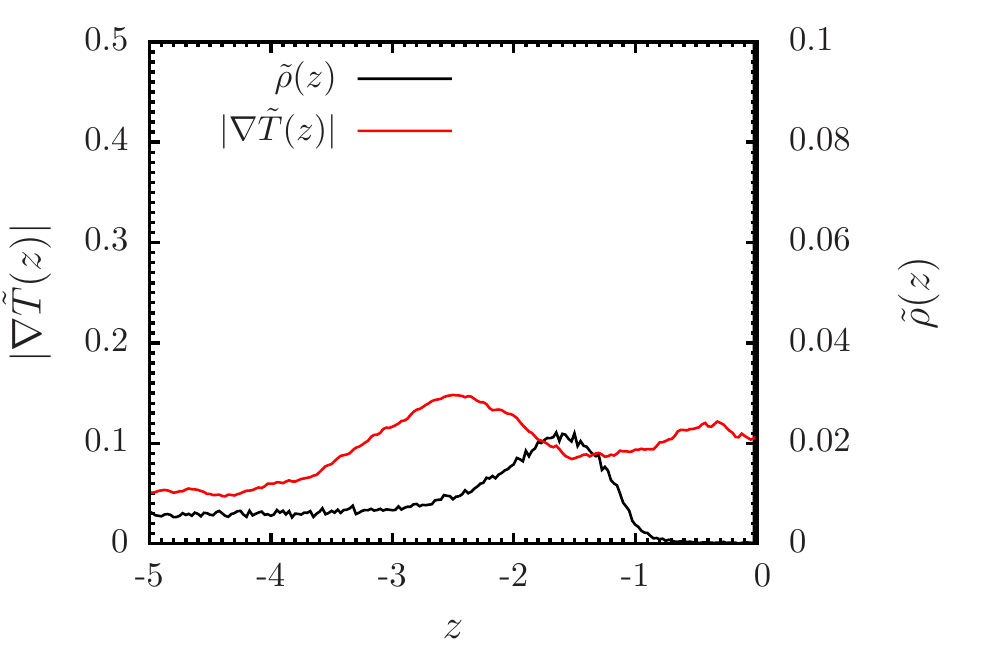}  \\ \includegraphics[width=0.5\linewidth]{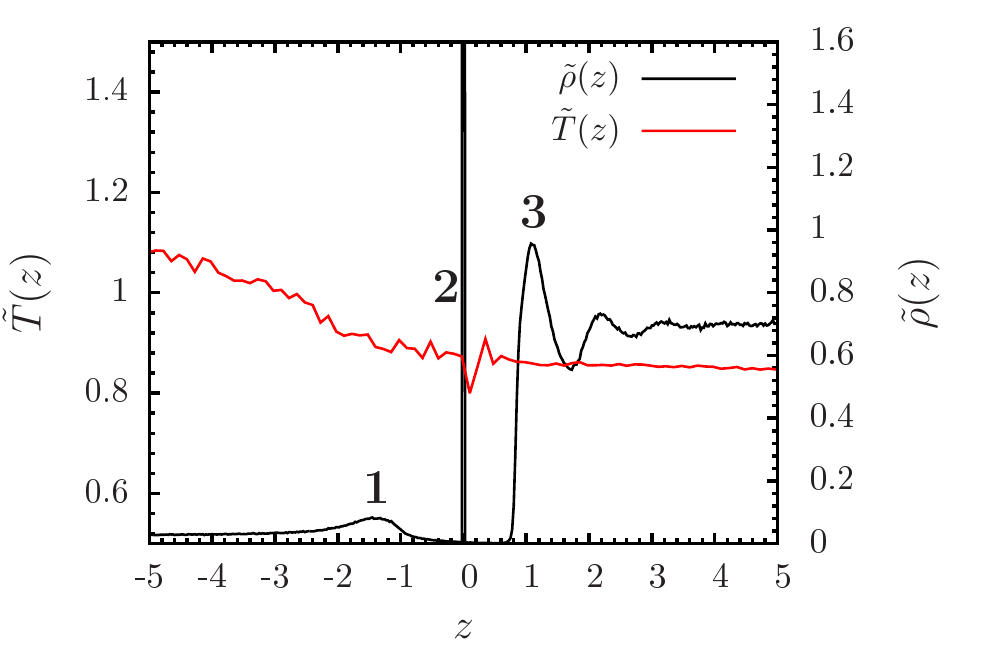}  &
 \includegraphics[width=0.5\linewidth]{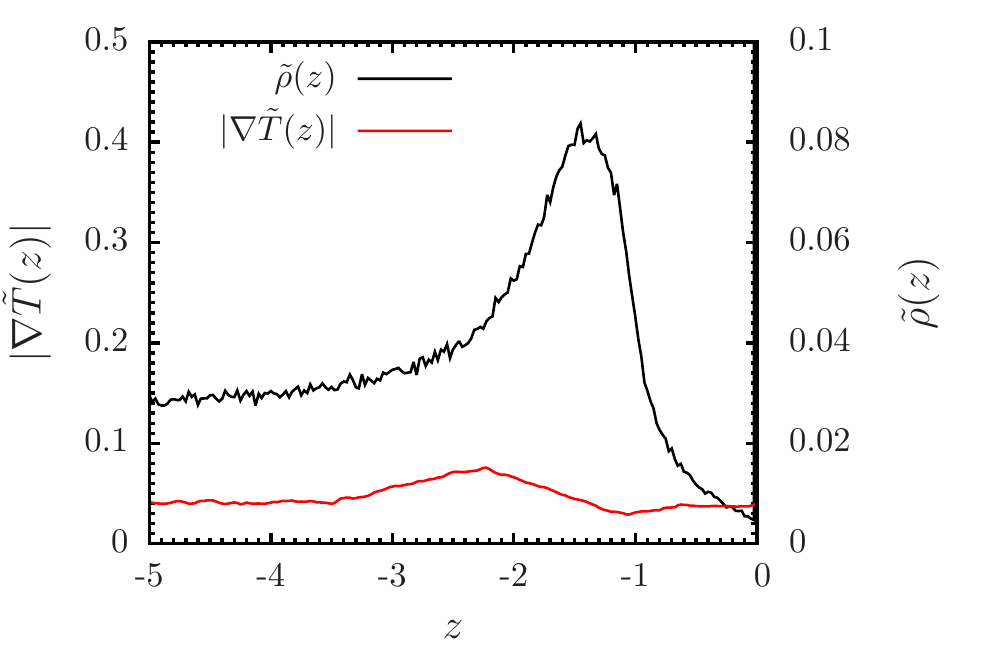} 
\end{tabular}
 \includegraphics[width=0.5\linewidth]{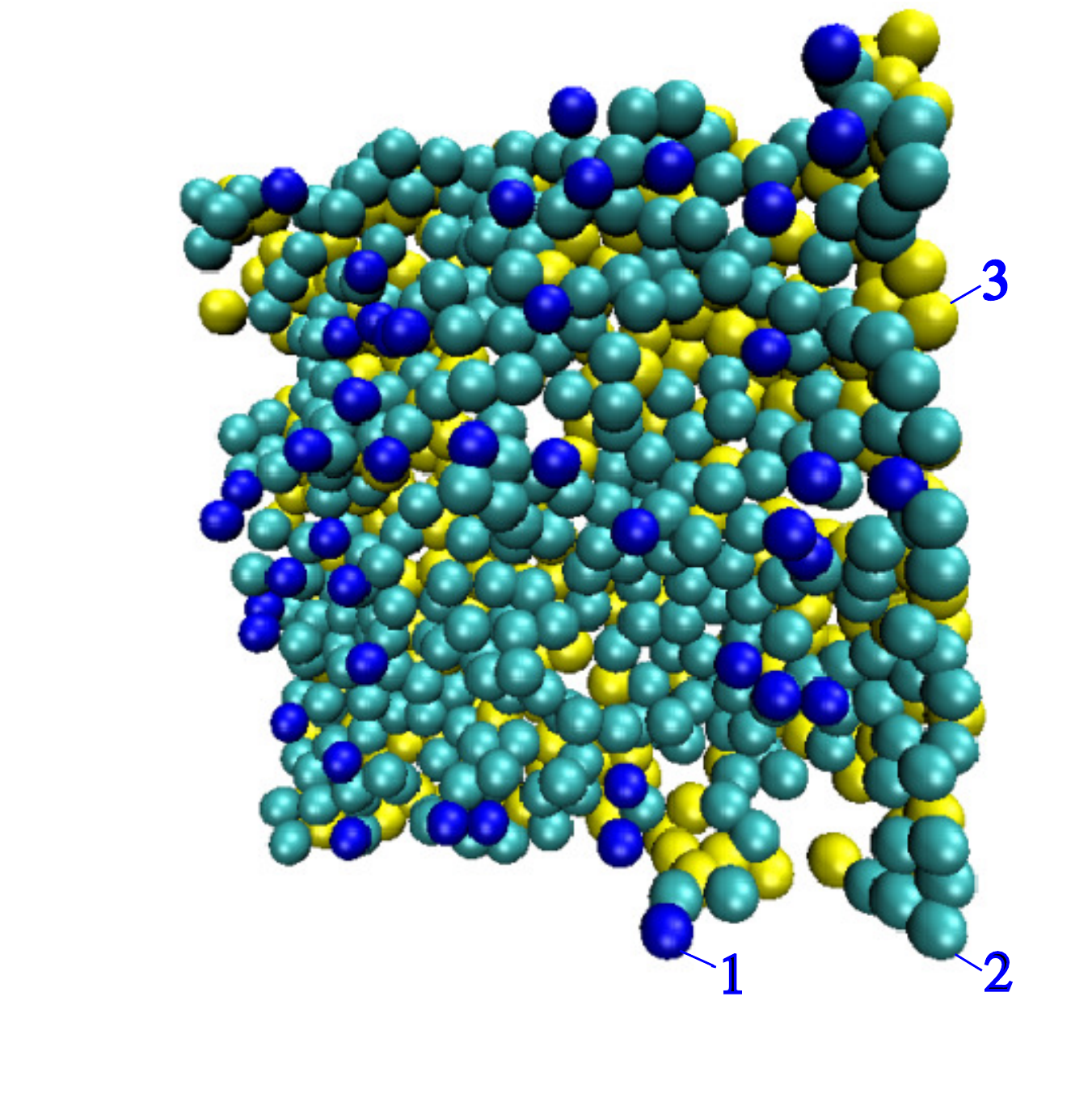} 
\caption{\label{fig:int_gradT_rho} Top-Left panels: Intrinsic temperature
profiles ($\tilde{T}(z)$, red line)  and
         intrinsic density profiles
         ($\tilde{\rho}(z)$, black line) for systems at different average temperatures, system I (top),
         system II (middle), and system~III 
		 (bottom). Top-Right panels: Absolute
         values of the gradient of the intrinsic temperature profiles ($|\nabla\tilde{T}(z)|$, red line) and 
         intrinsic density profiles ($\tilde{\rho}(z)$, black line) shifted and scaled to highlight the adsorbed layer
         for systems I
		 (top),
         II 
		 (middle) , and III 
		 (Bottom).
         The numerical derivative of the temperature profile is calculated from data
         averaged over a 0.2$\sigma$ window
         with the derivative further smoothed over a window of the same size.  
         The labels `1, 2, 3' refer to the fluid layers represented in the bottom panel. Bottom-central panel: Representative surface configuration for
system III. Atoms at the IS are represented in cyan. The atoms colored in blue represent the first layer
in the vapor phase, and the yellow atoms the first liquid layer. 
}
\end{figure}

%
The intrinsic profiles for different liquid temperatures 
are shown in Figure
\ref{fig:int_gradT_rho}. The delta function peaks in the density profiles (located at
$z=0.0$) are defined by the $N_s$ surface atoms at the IS. 
The oscillations in the liquid phase reflect the ordered layers of atoms in the liquid with respect to the
IS. 
It can also be seen that there is a low density peak 
in the vapor region. 
While small as compared to the
bulk liquid density, the peak is large in comparison to the bulk density of the vapor
(see right panels in Figure~2).
This peak is connected to the adsorption of atoms at the intrinsic surface.
In the percolation analysis performed in the
ISM, these atoms are identified as protrusions in 
the percolating cluster that defines the liquid surface and are connected to the 
bulk liquid, typically, by less than three neighbors. 
A visual representation of the configuration of particles corresponding to the peaks
either side of the intrinsic surface ($z=0.0$) illustrates this idea (see 
Figure~\ref{fig:int_gradT_rho}).
Particles colored in cyan represent the atoms belonging to
the $N_S$ 
surface sites in the IS. 
The particles colored in yellow represent the atoms in the interval
$0.7\leq z < 1.7$, {\it i.e.} the first liquid layer. The particles colored in blue
(interval $-2.7 \leq z < 0.0$) correspond to atoms in the layer
adsorbed at the IS. 
Effectively, the adsorption layer is similar to that formed by a gas adsorbing on a rough solid substrate.

The
thermal boundary resistance is located in the vapor phase,
above the liquid surface (see $\tilde{T}(z)$ and $|\nabla \tilde{T}(z)|$ in Figure~\ref{fig:int_gradT_rho}).
The temperature of the liquid surface 
(see Figure~2), is essentially the same as that of the bulk liquid. The thermal resistance, as given by
the temperature drop from the vapor to the liquid phase, depends strongly
on the average temperature of the run, and as expected, it is more prominent at low temperatures, indicating higher resistance to thermal transport. At
higher temperatures the
difference in temperature between the two bulk phases decreases significantly, as
exemplified by case III 
in Figure~\ref{fig:int_gradT_rho}.

At the molecular scale, the thermal resistance does not lead to 
a discontinuity in the temperature profile, but instead to a larger but continuous variation of the temperature gradient over a small length scale of a few atomic radii. 
This drop in 
temperature 
appears immediately above
the liquid surface. Figure~\ref{fig:int_gradT_rho} shows the absolute value of
the thermal gradient with reference to the 
intrinsic density profile in the vapor phase. 
The region of maximum thermal resistance 
coincides with the adsorption peak in the density of the vapor phase. 
As the density of the adsorbed layer decreases,
the temperature gradient, and therefore the thermal resistance, increases. We
propose that 
the adsorbed layer, the {\it thermal resistance layer} (TRL), 
controls the
exchange of heat between the liquid and vapor phases.

\begin{table}[!h]
\centering
 \begin{tabular}{cccccccc}
 \hline
  Interface & System & $L_x$/$\sigma$ & $L_z$ & $\dot{Q}$ & $\lambda_{V}$ &
$\lambda_{L}$ & $G_{K}$ \\
 \hline
  Liquid/Vapour & I & 22.2 & 1368.0 & 5 & 0.65 & 6.53 & 0.011  \\
  Liquid/Vapour & II & 22.2 & 1368.0 & 5 & 0.72 & 6.00 & 0.020  \\
  Liquid/Vapour & III & 20.5 & 256.5 & 10 & 0.55 & 5.00 & 0.082  \\
 \hline
 \end{tabular}
\caption{\label{tbl:thermalresistance} Summary of thermal transport properties
calculated for all systems simulated in this work, including the thermal
conductivities in the vapor phase ($\lambda_V$) and liquid phase
($\lambda_{L}$) and the interfacial thermal conductance ($G_K=J_q/\Delta T$).
The rate of energy input ($\dot{Q}$) is also listed along with the system
dimensions and initial temperature before a heat flux is applied. All quantities are in reduced
units.}
\end{table}

In order to calculate the thermal conductance, $G_K$, an 
extrapolation of $\tilde{T}(z)$ from each bulk phase is taken up to a precise `thermal boundary'  located at 
$z_K$, at which the temperature `jump', $\Delta T$, can be calculated.
This temperature `jump'
provides a nanoscopic description of thermal transport at the interface.
The obvious choice to define $z_K$ is given by the integral of the local
temperature deviation with respect to the temperature in the bulk gas, $T_g(z)$, and liquid, $T_l(z)$, phases, 
\begin{equation}
\int_{z_g}^{z_K} (\tilde{T}(z)-T_g(z))dz +
\int_{z_K}^{z_l} (\tilde{T}(z)-T_l(z))dz=0,
\end{equation}
where the positions $z_g$ and $z_l$ are located well within the respective
asymptotic regimes $\tilde{T}(z)=T_g(z)$ for $z<z_g$ and $\tilde{T}(z)=T_l(z)$ for $z>z_l$. An
illustration of how we identify $z_K$ is given in the SI. The approach introduced here provides a unique definition of the
temperature `jump', as there is no need to define an arbitrary location for the extrapolation of the temperature profiles in order to calculate $\Delta T$. 

The values of the interfacial thermal resistance for each average temperature are reported in Table~\ref{tbl:thermalresistance} along with the
thermal conductivities of the respective bulk phases. The thermal
conductance increases with increasing average temperature,
changing in magnitude by a factor of eight from system I to system III. 
This increase is
accompanied by an enhancement of the thermal conductivity and density in the gas phase.
The thermal conductivities in the liquid phase
are of the same order as the experimental values for argon~\cite{nist}. 
The simulated thermal conductivity 
for
system II (temperature of the liquid $\sim$84~K) is $\lambda_L=0.11\text{ Wm}^{-1}\text{K}^{-1}$, in good agreement with experimental data, $\lambda_L=0.13\text{ Wm}^{-1}\text{K}^{-1}$. 

\begin{figure}
 \includegraphics[width=1\linewidth]{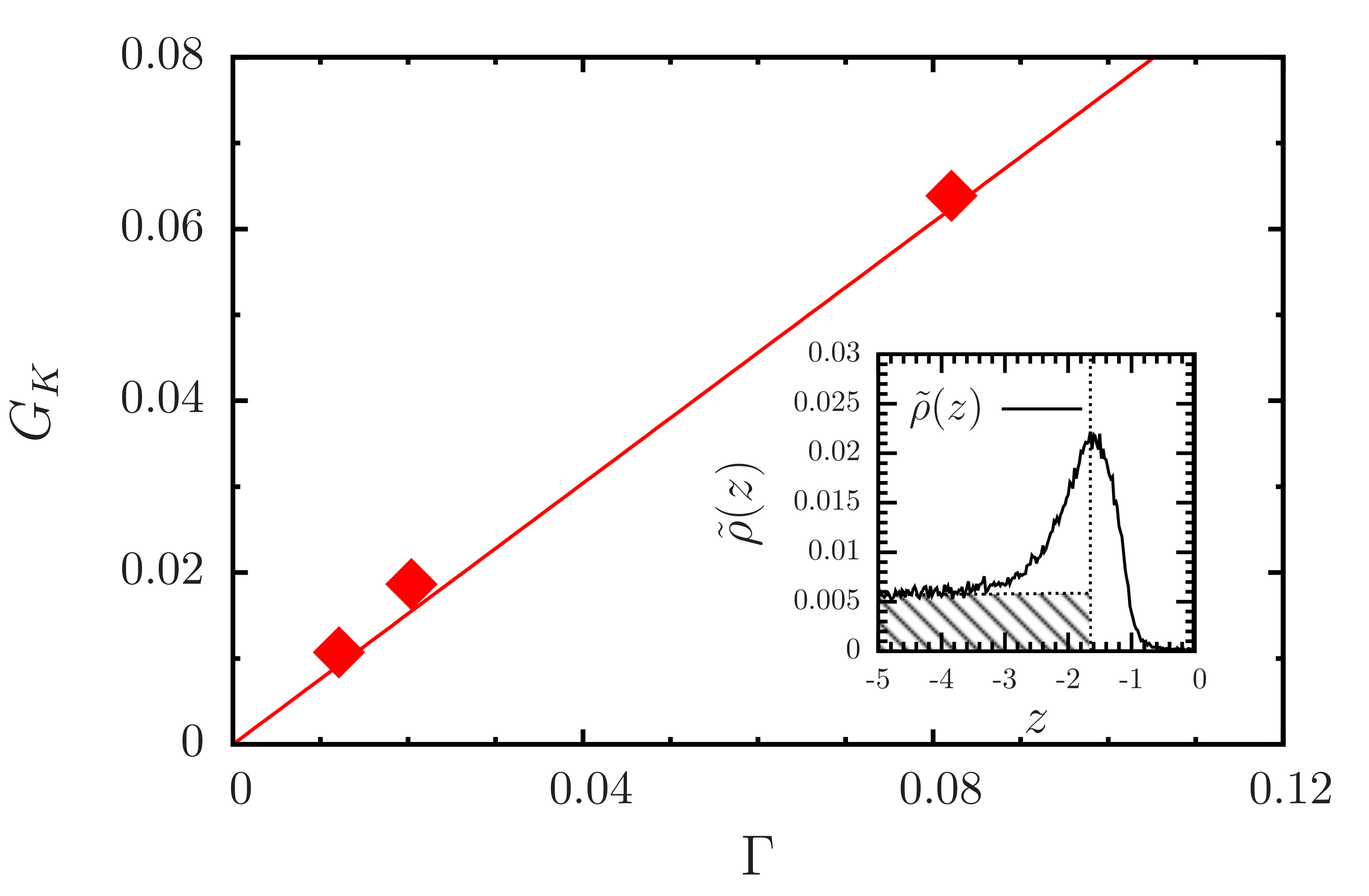}
 \caption{\label{fig:G_adsrb}Thermal conductance $G_K$ against adsorption
 $\Gamma$. The red line is
 a guide for the eye. The inset plot illustrates the calculation of the adsorption
 from equation \eqref{eq:adsorption} for system~II. 
 The shaded region is subtracted from the integral over the intrinsic
 density profile.}
\end{figure}

We 
compute now the fluid
adsorption
associated to the TRL. 
We define the adsorption as
\begin{equation}\label{eq:adsorption}
 \Gamma = \frac{1}{m}\int_{-\infty}^0\tilde{\rho}(z)dz -
 \frac{1}{m}\int_{-\infty}^{z_{\text{max}}}\rho_{g}(z)dz
\end{equation}
where $m$ is the mass of the particle ($m=1$ in reduced units), $\rho_{g}(z)$
is a linear extrapolation of the density profile in the vapor phase at
sufficient distance from the interface and $z_{\text{max}}$ is the
position of the local maximum due to adsorbed particles. 
Figure~\ref{fig:G_adsrb} shows
the thermal conductance $G_K$ plot as a function of the surface
adsorption $\Gamma$ for the systems I, II, and III discussed above.
A linear relationship is observed 
between the conductance and the
adsorption, supporting the notion that the thermal resistance is governed
by the degree of adsorption in the TRL that separates the
liquid surface from the vapor phase.





We have shown that the combination of non-equilibrium molecular dynamics simulations and intrinsic sampling approaches provides a route to identify the interfacial structure defining the thermal resistance of liquid-vapor interfaces. We draw the following conclusions from our study:

$\bullet$ The thermal conductance of liquid-vapour interfaces is defined by the 
{\it thermal resistance layer}, which is located at $\sim 2$~molecular diameters from the mean position of the liquid surface and shifted towards the vapor phase. The TRL is a boundary resistance layer for the transition from ballistic thermal transport (vapor), to thermal transport dominated by atomic/molecular collisions inside the cage formed by the nearest neighbors (liquid). 

$\bullet$ The liquid surface (defined by the intrinsic surface) and the bulk liquid have essentially the same temperature.

$\bullet$ In the sharpest representation of the interface, obtained with the ISM, the TRL is defined by a local maximum in the density 
in the vapor region. 
The TRL is defined by 
the adsorption of a low density monolayer at the liquid surface, which within the ISM percolation
analysis correspond to atom `overhangs' of 
the liquid phase. 
The 
transition in the heat transfer mechanism between the vapor 
and the liquid phases, proceeds via
the atomic overhangs, which form via a dynamical equilibrium where molecules from the 
hot gas get trapped after a collision with the liquid surface and molecules from the cold liquid
reach the surface and eventually detach from it. The sharp temperature change $\Delta T$ at the TRL reflects
precisely the energy transfer between these two distinct types of heat carriers.  

$\bullet$ 
The density of the TRL 
increases as the critical point is approached, 
{\it i.e.} with the density of the coexisting vapor. The thermal conductance increases linearly with the number of particles in the TRL. 

We propose that 
the TRL is the key molecular structure determining the thermal transport
across liquid-vapor interfaces and that the number of heat carriers in the TRL may be quantified in computer 
simulations when the blurring effects of the capillary waves are eliminated from the temperature profile $T(z)$.
Further work 
will focus on the investigation of  
the thermal resistance of multicomponent systems.

We acknowledge the EPSRC-UK (EP/J003859/1) and the Spanish Secretariat for Research, Development and Innovation (Grant No. FIS2013-47350-C5 and MDM-2014-0377) for financial support.
We thank the Imperial College High Performance Computing Service for providing computational resources and Dick Bedeaux for insightful discussions on heat transport across interfaces.


%

\end{document}